\documentclass[conference, 10pt]{IEEEtran}

\usepackage{cite}
\usepackage[cmex10]{amsmath}
\usepackage{url}
\usepackage{graphicx}
\usepackage{color}
\usepackage{placeins}
\usepackage{float}
\usepackage{tabularx,colortbl}
\usepackage{ifthen}

\makeatletter

\newcounter{author}
\renewcommand{\author}[2][]{
   \stepcounter{author}
   \@namedef{author@\theauthor}{#2}
   \@namedef{authorlabel@\theauthor}{#1}
}

\newcounter{address}
\newcommand{\address}[2][]{
   \stepcounter{address}
   \@namedef{address@\theaddress}{#2}
   \@namedef{addresslabel@\theaddress}{#1}
}

\newcommand{\alsep}{and}

\def\newmaketitle{\par%
  \begingroup%
  \normalfont%
  \def\thefootnote{}%  the \thanks{} mark type is empty
  \def\footnotemark{}% and kill space from \thanks within author
  \let\@makefnmark\relax% V1.7, must *really* kill footnotemark to remove all \textsuperscript spacing as well.
  \footnotesize%       equal spacing between thanks lines
  \footnotesep 0.7\baselineskip%see global setting of \footnotesep for more info
  \normalsize%
  \twocolumn[\thenewmaketitle\@IEEEaftertitletext]%
  \if@IEEEusingpubid
     \enlargethispage{-\@IEEEpubidpullup}%
  \fi
  \endgroup
  \setcounter{footnote}{0}\let\maketitle\relax\let\@maketitle\relax
  \gdef\@thanks{}%
  \let\thanks\relax}

\def\thenewmaketitle{
  \newpage
  \begin{center}%
    \vskip0.2em{\Huge\@IEEEcompsoconly{\sffamily}\@IEEEcompsocconfonly{\normalfont\normalsize\vskip 2\@IEEEnormalsizeunitybaselineskip
   \bfseries\large}\@title\par}\vskip1.0em\par%
    \vspace{1ex}
    \newcounter{c@author}
    \newcounter{c@tmp}
    \ifthenelse{\value{author}=2}{%
      \newcommand{\liand}{ and }}{%
      \newcommand{\liand}{, and }}
    \ifthenelse{\value{address}<2}{%
      \@nameuse{author@1}%
      \stepcounter{c@author}%
      \whiledo{\value{c@author}<\value{author}}{%
        \setcounter{c@tmp}{\value{author}}%
        \addtocounter{c@tmp}{-\value{c@author}}%
        \ifthenelse{\value{c@tmp}=1}{%
          \renewcommand{\alsep}{\liand}}{\renewcommand{\alsep}{, }}%
        \stepcounter{c@author}\alsep \@nameuse{author@\thec@author}}\\%
    }
    {%Add address references after the author's name
      \@nameuse{author@1}${}^{(\ref{\@nameuse{authorlabel@1}})}$%
      \stepcounter{c@author}%
      \whiledo{\value{c@author}<\value{author}}{%
      \setcounter{c@tmp}{\value{author}}%
      \addtocounter{c@tmp}{-\value{c@author}}%
      \ifthenelse{\value{c@tmp}=1}{%
        \renewcommand{\alsep}{\liand}}{\renewcommand{\alsep}{, }}%
      \stepcounter{c@author}\alsep \@nameuse{author@\thec@author}%
        ${}^{(\ref{\@nameuse{authorlabel@\thec@author}})}$%
      }
    }
    \vspace{0.2ex}

    \ifthenelse{\value{address}>0}{%
      \ifthenelse{\value{address}=1}{
        {\@nameuse{address@1}}
      }
      {%Output the addresses as an enumerated list
        \newcounter{c@address}

        \begin{center}
        \whiledo{\value{c@address}<\value{address}}
        {
          \refstepcounter{c@address}
            ${}^{(\thec@address)}$\,%
              \label{\@nameuse{addresslabel@\thec@address}}%
              \@nameuse{address@\thec@address}\\ %
        }
        \end{center}
      } % end of the address creation ifthenelse block
    }
    {
      \relax
    }
  \end{center}
}

\makeatother

\title{Non-Local Metasurface-aided Leaky-Wave Antennas}

\author[org1]{Seokjun Kim}
\author[org1]{Haneul Ryu}
\author[org1]{Jeong-Hae Lee}
\author[org1]{Minseok Kim\textsuperscript{*}}

\address[org1]{School of Electronic and Electrical Engineering, Hongik University, 94 Wausan-ro, Mapo-gu, Seoul, 121-791, Korea \\ \textsuperscript{*}minseok.kim@hongik.ac.kr}
\begin{document}

\newmaketitle

\begin{abstract}
This work presents a non-local terahertz metasurface integrated into a leaky-wave antenna for robust, wide-angle beam steering. The metasurface encodes a holographic pattern by explicitly inducing tangential and normal susceptibilities, along with magnetoelectric coupling. This design maintains stable radiation performance even when the longitudinal wavenumber of the incident guided mode—and thus its effective impinging angle—varies as a function of frequency. In particular, we show that there exists a limit to achieving exact angular insensitivity and propose an optimization-based framework to obtain the required susceptibilities that closely approximate near angle-insensitive performance for stable beam-steering performance. Additionally, an iterative synthesis approach is introduced that maps abstract susceptibilities to physically realizable structures. Full-wave simulations demonstrate a beam-scanning range of nearly 50$^\circ$ over the 2.0–2.7 THz band—a more than threefold improvement over conventional local-metasurface designs.
\end{abstract}

\section{Introduction}
Recently, metasurface-aided leaky-wave antennas (LWAs) have emerged as a promising solution to the physical constraints of traditional LWA designs. By employing metasurfaces as radiating apertures, these antennas have been shown to overcome the open-stopband effect~\cite{Liu2002ElecLett}, enable independent control over the leakage factor and phase constant~\cite{Abdo2019IEEE}, and facilitate the synthesis of complex far-field patterns, such as sector and Dolph-Chebyshev beams~\cite{Kim2021AWPL}. However, this potential is often limited by the conventional practice of synthesizing the response from a strictly \textit{local} perspective, assuming a fixed angle of incidence. In practice, the frequency-scanning mechanism of an LWA inherently modifies the effective incidence angle. When this angle shifts from the design point, the metasurface fails to sustain a consistent scattering performance, leading to substantial gain reduction across the operational band.

In contrast, non-local or spatially dispersive metasurfaces operate on a fundamentally different principle. Non-locality suggests that electromagnetic boundary conditions are sensitive not just to local field strength, but to its spatial distribution and gradients. In the context of generalized sheet transition conditions (GSTCs), non-locality is often represented by coupling spatial gradients ($\nabla_t$) to the normal components of the polarization densities~\cite{Shaham2025IEEETAP}. Since these spatial derivatives map directly to transverse wavevectors in the spectral domain—and consequently to the angle of incidence—the inclusion of normal polarization provides the angular selectivity necessary to distinguish between waves impinging from different directions.

In this work, we leverage normal susceptibility to ensure angular stability in metasurface-aided LWAs, thereby overcoming the gain degradation typically encountered in local designs. Given the growing interest in the terahertz (THz) regime for next-generation applications, the proposed non-local metasurface-aided LWA is designed to operate near 2.5 THz with the goal of achieving angle-insensitive behavior within the metasurface. Notably, full-wave simulations verify a beam-steering range of nearly $50^\circ$ through broadside, successfully overcoming the open-stopband effect and demonstrating a more than threefold expansion in scanning coverage compared to local counterparts.

\section{Non-local metasurface design framework}
\begin{figure}[t!]
\begin{center}
\noindent
  \includegraphics[width=\linewidth]{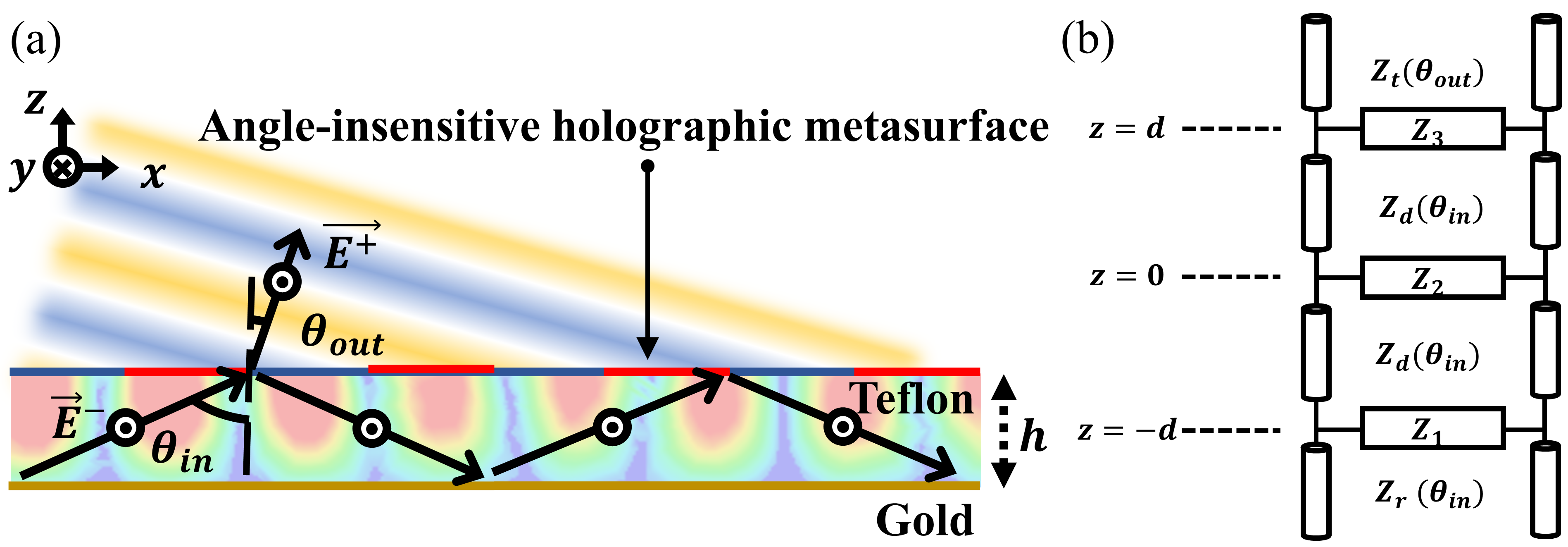}
  \caption{(a) Schematic of the proposed leaky-wave antenna and (b) the equivalent transmission-line model of the non-local metasurface.}\label{fig:Str_LWA}
\end{center}
\vspace{-0.65cm}
\end{figure}

The overall architecture of the THz LWA, incorporating the proposed non-local metasurface, is shown in Fig.~\ref{fig:Str_LWA}.
The antenna consists of a 300~nm gold (Au) ground plane, a 50.8~$\mu$m Teflon substrate, and the metasurface comprising three cascaded Au-patterned impedance layers separated by 5~$\mu$m polyimide spacers. This architecture is specifically chosen, as it supports both tangential and normal surface susceptibilities~\cite{Shaham2025IEEETAP}. Under TE polarization, reciprocity, and losslessness, the structure is fully described by four susceptibilities: tangential electric ($\chi_{ee}^{yy}$) and magnetic ($\chi_{mm}^{xx}$), magnetoelectric coupling ($\chi_{em}^{yx}$), and normal magnetic ($\chi_{mm}^{zz}$). Here, $\chi_{ee}^{yy}$, $\chi_{mm}^{xx}$, and $\chi_{mm}^{zz}$ are real-valued, while $\chi_{em}^{yx}$ is purely imaginary, each linking a surface polarization to its macroscopic field.

To achieve angle-independent behavior, we first consider a uniform metasurface composed of identical unit cells and characterize the angular scattering response by solving the GSTCs~\cite{Budhu2021Nanophotonics}. This leads to closed-form expressions for the reflection and transmission coefficients as rational functions of $\tilde{k}_{z,t} = \cos(\theta_t)$ and $\tilde{k}_{z,r} = \sqrt{\epsilon_r} \cos(\theta_i)$:
%For a uniform metasurface on a substrate of permittivity $\epsilon_r$, solving the GSTCs yields the angular scattering response~\cite{Budhu2021Nanophotonics}. This leads to closed-form expressions for the reflection and transmission coefficients as rational functions of $\tilde{k}_{z,t} = \cos(\theta_t)$ and $\tilde{k}_{z,r} = \sqrt{\epsilon_r} \cos(\theta_i)$:
\begin{subequations}
\label{eq:long_eq_bottom}
\begin{align}
r_{\text{zt}}(\theta_i) &=\frac{R_\text{poly}}{D_\text{poly}},
\label{eq:r_zt}
\\[8pt]
t_{\text{zt}}(\theta_i) &=\frac{T_\text{poly}}{D_\text{poly}},
\label{eq:t_zt}
\end{align}
\label{eq:rt_zt}
\end{subequations}
where the polynomial fractional terms are
{\allowdisplaybreaks
\begin{subequations}
\begin{align}
R_\text{poly} =&~r_0 + r_1^t \tilde{k}_{z,t} + r_1^r \tilde{k}_{z,r}
+ r_2^t \tilde{k}_{z,r}\tilde{k}_{z,t}
+ r_2^r \tilde{k}_{z,r}^2\\\notag
&+ r_3^t \tilde{k}_{z,r}^2\tilde{k}_{z,t}
+ r_3^r \tilde{k}_{z,r}^3, \\
D_\text{poly} =&~d_0 + d_1^t \tilde{k}_{z,t} + d_1^r \tilde{k}_{z,r}
+ d_2^t \tilde{k}_{z,r}\tilde{k}_{z,t}
+ d_2^r \tilde{k}_{z,r}^2\\\notag
&+ d_3^t \tilde{k}_{z,r}^2\tilde{k}_{z,t}
+ d_3^r \tilde{k}_{z,r}^3, \\
T_\text{poly} =&~t_1 \tilde{k}_{z,r} + t_3 \tilde{k}_{z,r}^3,
\end{align}
\end{subequations}}
with the coefficients further defined as
{\allowdisplaybreaks
\begin{subequations}
    \label{eq:rtd_coefficientsr}
    \begin{align} 
    &r_0=-j4(\tilde{\chi}_{ee}^{yy}+\tilde{\chi}_{mm}^{zz}),\\
    &r_1^t = -(j \tilde{\chi}_{em}^{yx}-2)^2 + (\tilde{\chi}_{ee}^{yy}+\tilde{\chi}_{mm}^{zz})\tilde{\chi}_{mm}^{xx},\\
    &r_1^r = (j \tilde{\chi}_{em}^{yx}+2)^2-(\tilde{\chi}_{ee}^{yy}+\tilde{\chi}_{mm}^{zz})\tilde{\chi}_{mm}^{xx},\\
    &r_2^t=j4\tilde{\chi}_{mm}^{xx}, \\
    &r_2^r = j4\tilde{\chi}_{mm}^{zz},\\
    &r_3^t=-\tilde{\chi}_{mm}^{zz}\tilde{\chi}_{mm}^{xx},\\
    &r_3^r=\tilde{\chi}_{mm}^{xx}\tilde{\chi}_{mm}^{zz},\\
    &t_1=8+2(\tilde{\chi}_{em}^{yx})^2+2(\tilde{\chi}_{ee}^{yy}+\tilde{\chi}_{mm}^{zz})\tilde{\chi}_{mm}^{xx}, \\
    &t_3=-2\tilde{\chi}_{mm}^{zz}\tilde{\chi}_{mm}^{xx},\\
    &d_0=j4(\tilde{\chi}_{ee}^{yy}+\tilde{\chi}_{mm}^{zz})\\
    &d_1^t=(j \tilde{\chi}_{em}^{yx}-2)^2-(\tilde{\chi}_{ee}^{yy}+\tilde{\chi}_{mm}^{zz})\tilde{\chi}_{mm}^{xx},\\
    &d_1^r=(j \tilde{\chi}_{em}^{yx}+2)^2-(\tilde{\chi}_{ee}^{yy}+\tilde{\chi}_{mm}^{zz})\tilde{\chi}_{mm}^{xx},\\
    &d_2^t=j4\tilde{\chi}_{mm}^{xx}, \\
    &d_2^r=-j4\tilde{\chi}_{mm}^{zz},\\
    &d_3^t=\tilde{\chi}_{mm}^{zz}\tilde{\chi}_{mm}^{xx}, \\
    &d_3^r=\tilde{\chi}_{mm}^{xx}\tilde{\chi}_{mm}^{zz}.
    \end{align}
\end{subequations} 
}
%\noindent Here, $\theta_i$ and $\theta_t$ denote the incident and transmitted angles for a uniform metasurface on a substrate of permittivity $\epsilon_r$.
\noindent Here, $\theta_i$ and $\theta_t$ denote the incident and transmitted angles.
With the angular scattering response obtained in ~\eqref{eq:rt_zt}, the surface susceptibilities are related to the multilayer impedances $\{Z_1, Z_2, Z_3\}$ based on the transmission-line theory~\cite{Shaham2025IEEETAP}. This establishes a complete link between the angular response, the surface susceptibilities, and the physical impedance parameters. We then examine the extent to which the four surface susceptibilities contribute to angular insensitivity by imposing the condition $r_\text{zt}(\theta_i)=\bar{R}$, where $\bar{R}$ is a prescribed, angle-independent reflection coefficient.

To enforce $r_\text{zt}(\theta_i) = \bar{R}$ for all incidence angles, the angle-dependent terms in~\eqref{eq:rt_zt} must cancel identically, leading to:
\begin{subequations}
\begin{align}
&r_n^{r,t} = \Re\{\bar{R}\} d_n^{r,t}, \label{eq:cond1} \\
&j \Im\{\bar{R}\} d_n^{r,t} = 0, \label{eq:cond2}
\end{align}
\end{subequations}
for $n \in \{0, 1, 2, 3\}$. When the media above and below the metasurface differ ($\tilde{k}_{z,t} \neq \tilde{k}_{z,r}$), it is seen that these constraints cannot be simultaneously satisfied unless $\bar{R} = \pm 1$. This implies that exact angular insensitivity is restricted to perfect electric ($\bar{R}=1$) or magnetic ($\bar{R}=-1$) conductors. However, while exact angle-invariant partial reflection ($|\bar{R}| < 1$) is analytically unattainable within this susceptibility model, we demonstrate that a near-insensitive response can be realized through numerical optimization.

\begin{figure}[t]
	\centering
 	\resizebox{0.45\textwidth}{!}{\includegraphics{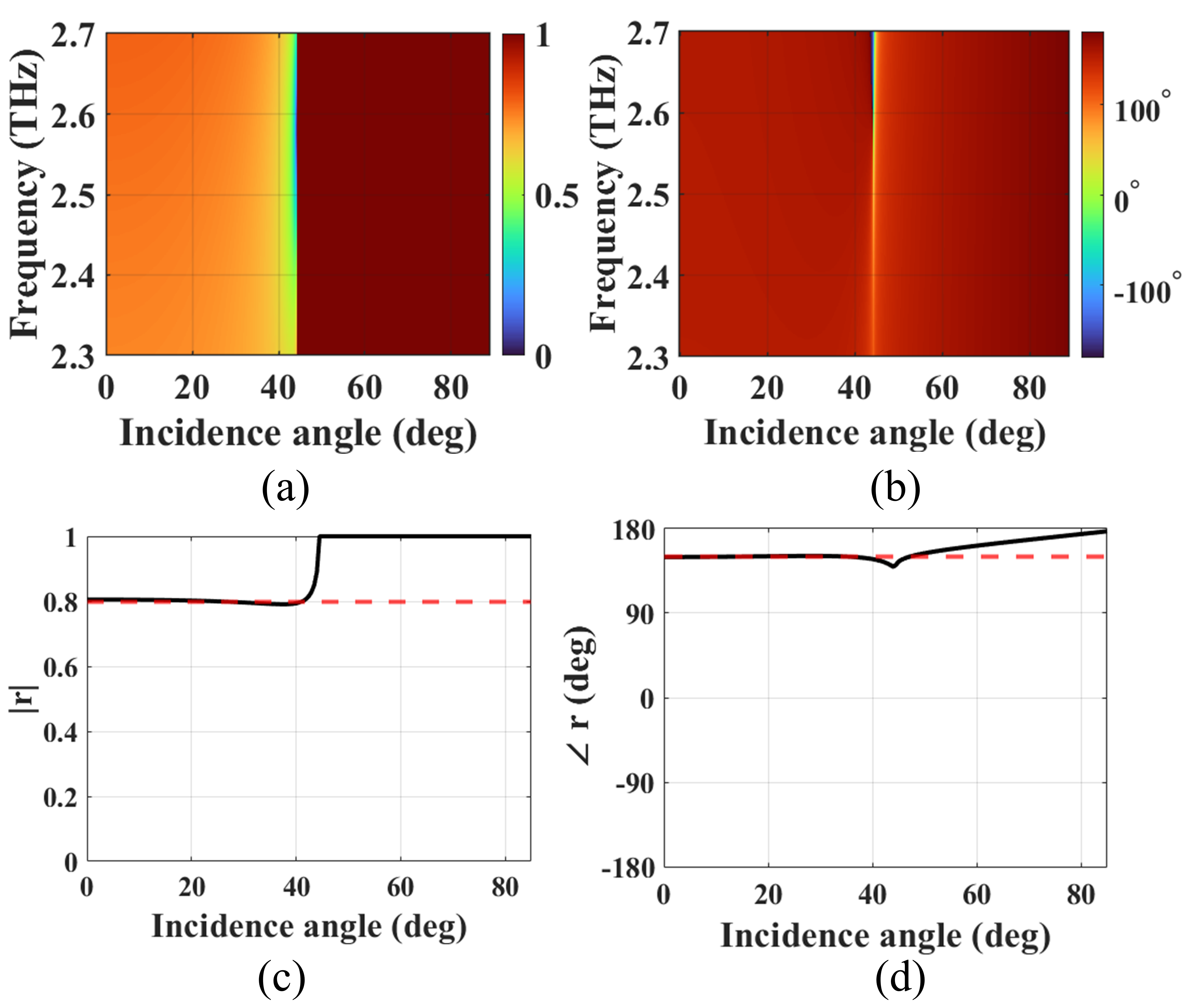}}
	\caption{Angular and frequency dependence of the optimized unit cell. (a) Magnitude and (b) phase of the reflection coefficient as a function of incidence angle and frequency. (c, d) Comparison of the simulated reflection (c) magnitude and (d) phase against the target values at 2.5 THz.}
	\label{fig:OptSingleUC}
\vspace{-0.35cm}
\end{figure}

As a representative result, Fig.~\ref{fig:OptSingleUC} shows the optimized reflection profile for a target value of $\bar{R} = 0.8e^{j150^\circ}$.
Here, various combinations of surface impedance values $\{Z_1, Z_2, Z_3\}$ are explored within a particle swarm optimization (PSO) routine to approximate the target angle-independent reflection coefficient. Based on this scheme, the optimized impedance values are determined to be $\{Z_1 = j102.2$, $Z_2 = j181.8$, and $Z_3 = -j270.7~\}~\Omega$. The resulting angular response maintains target performance up to the total internal reflection (TIR) onset at $\theta_\text{i} \approx 43^\circ$. Therefore, if this metasurface were directly integrated into the waveguide, it would enable stable beam steering over the internal propagation angle range $0<\theta_\text{in} < 43^\circ$. However, operation near broadside ($\theta_\text{in} \approx 0^\circ$) is hindered by the waveguide cutoff—leading to the open-stopband effect—and increased dielectric losses from extended propagation paths. 

To bypass TIR constraints, we adopt a holographic approach where a transmission profile encodes the interference between guided and radiative waves~\cite{Kim2021PRApplied}. However, given the analytical complexity of globally non-uniform, non-local surfaces, the design employs a local periodic approximation: the macroscopic layout follows periodic assumptions using discrete sampling elements, while each microscopic unit cell is engineered for spatial dispersion to suppress angular sensitivity. In other words, the non-local behavior is intentionally embedded within individual cells to suppress angular sensitivity, but coupling between neighboring non-identical cells is neglected.

Building on this framework, two binary unit cells (Cell 1 and Cell 2) are optimized via PSO to encode the holographic pattern $\tau(x)$ defined by
\begin{equation}
\tau(x) = (1 - |\bar{R}|^2) \mathrm{sgn}\left(\cos\left(k_0 \left|\sin\theta_\text{in} - \sin\theta_\text{out}\right| x\right)\right).
\label{eq:HoloPatt}
\end{equation}
To overcome the open-stopband effect and ensure propagation beyond the critical angle, both cells are designed for a shared angle-insensitive reflection coefficient of f$\bar{R} = 0.8e^{j38^\circ}$ at 2.5 THz, with $\theta_\text{in} = 60^\circ$ and a radiation angle $\theta_\text{out} = 15^\circ$. While their reflection properties are identical, their transmission phases are set to differ by 180$^\circ$ to enable binary sampling of the interference pattern. In this process, however, the longitudinal wavenumber $\tilde{k}_{z,t}$ in ~\eqref{eq:rt_zt} is modified to represent the imposed holographic pattern as $\tilde{k}_{z,t} = -j\sqrt{\left(\frac{\lambda_0}{2\pi}\frac{d\Phi}{dx} + \sqrt{n_i^2 - \tilde{k}_{z,i}^2}\right)^2 - 1}$
such that the phase gradient $d\Phi/dx$ is set to align with the macroscopic periodicity of the holographic pattern ($\Lambda_x$) (i.e., $d\Phi/dx \approx 2\pi/\Lambda_x$)

\begin{figure}[t!]
	\centering
	\resizebox{0.45\textwidth}{!}{\includegraphics{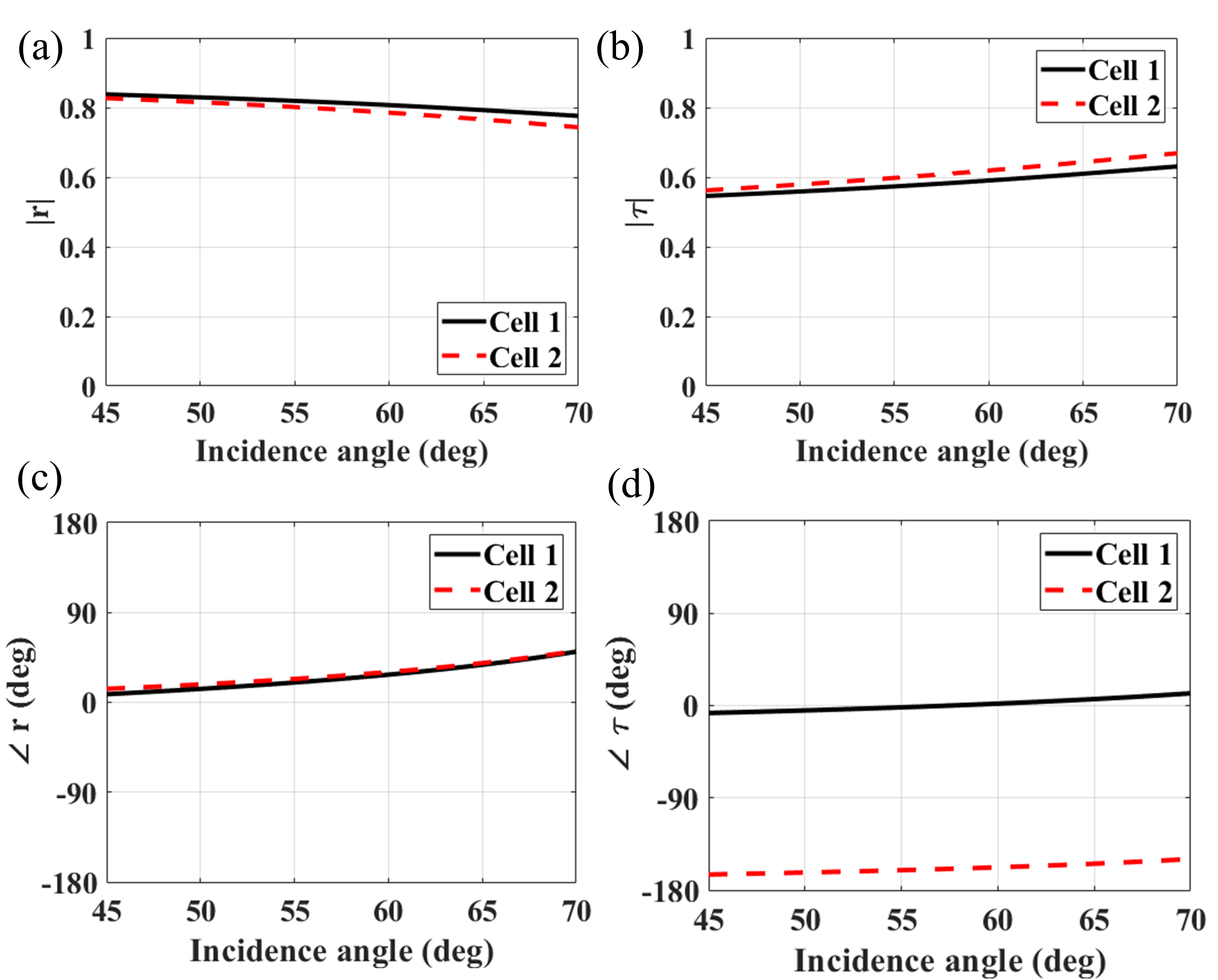}}
	\caption{Comparison of the scattering parameters for Cell 1 and Cell 2 at the design frequency 2.5 THz: (a) reflection magnitude, (b) transmission magnitude, (c) reflection phase,
and (d) transmission phase.}
	\label{fig:AngRes}
    \vspace{-0.5cm}
\end{figure}

Fig.~\ref{fig:AngRes} presents the angular responses of the optimized Cell~1 and Cell~2 at 2.5 THz, with surface impedances $\{Z_1 = -j253.7,~Z_2 = j144.5,~Z_3 = j119.8\}~\Omega$ and $\{Z_1 = -j176.8,~Z_2 = j500,~Z_3 = -j48\}~\Omega$, respectively.
As shown, both reflection and transmission magnitudes and phases remain stable across incidence angles from $45^\circ$ to $70^\circ$. %Nevertheless, perfect angular insensitivity is not achieved, as expected from the theoretical limitation that only PEC or PMC boundaries yield exact invariance under asymmetric longitudinal wavenumbers. Simultaneous stability of both reflection and transmission coefficients would require higher-order multipoles or angle-dispersive susceptibilities, which are challenging to physically implement, and it is beyond the scope of this work. The present design instead focuses on four angle-independent surface susceptibilities, practically realized via cascaded impedance layers.
Nevertheless, perfect angular insensitivity is not achieved, as expected from the theoretical limitation shown previously. Such surfaces would require a greater degree of nonlocality, which is challenging to physically implement and is beyond the scope of this work. The present design instead focuses on four angle-independent surface susceptibilities, practically realized via cascaded impedance layers.

\begin{figure}[t!]
	\centering
	\resizebox{0.4\textwidth}{!}{\includegraphics{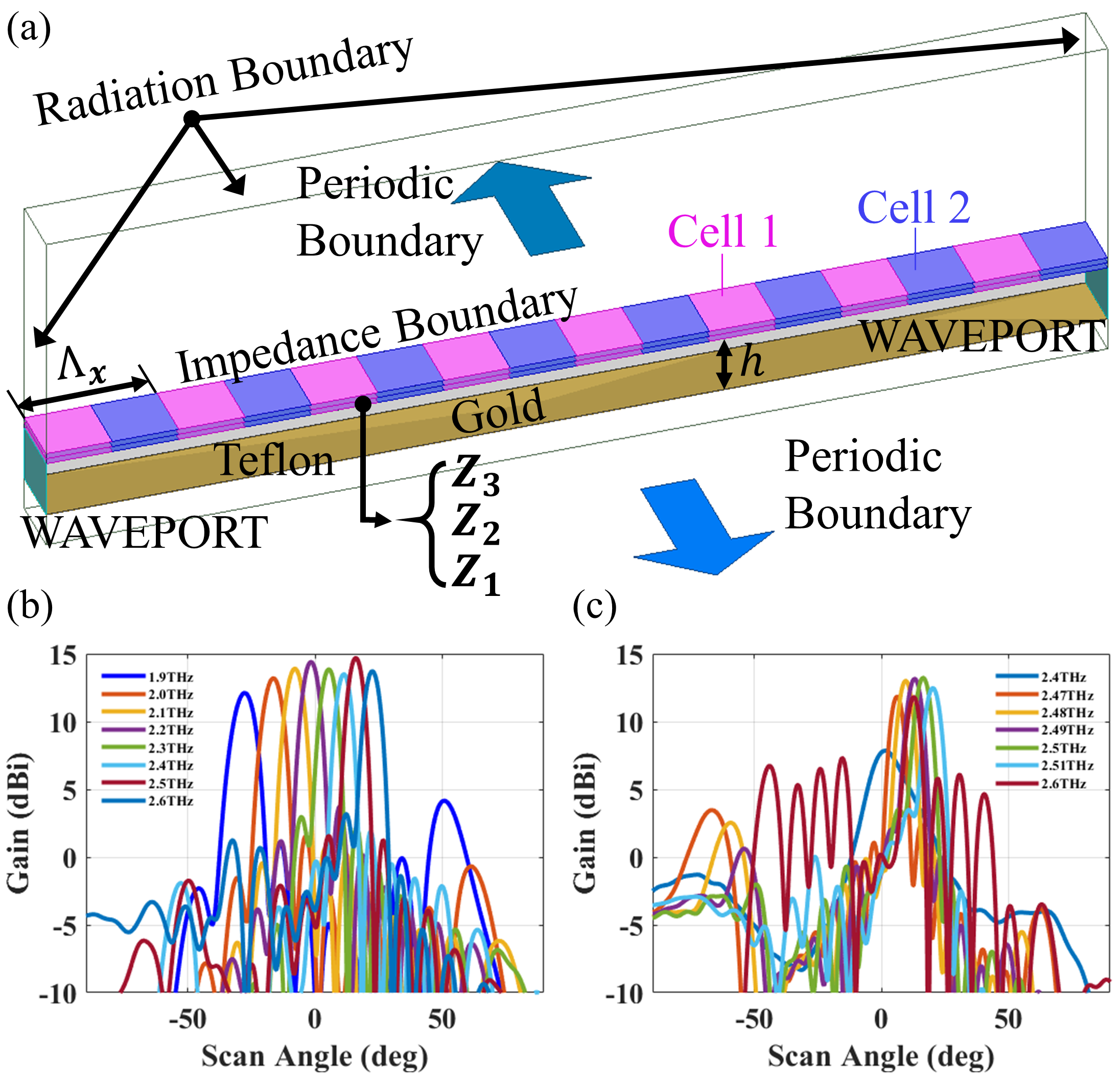}}
	\caption{(a) Schematic of simulation setup employing impedance boundaries that implement the optimized $\{Z_1,~Z_2,~Z_3\}$ for Cell~1 and Cell~2. Full-wave radiation patterns (gain) of the LWA employing (b) the proposed angle-insensitive metasurface over 1.9-2.6 THz and (c) local metasurface over 2.4-2.6 THz.}
	\label{fig:impedance layer 1D lwa}
    \vspace{-0.5cm}
\end{figure}

Despite the inherent limitations, the optimized metasurface offers sufficiently stable scattering responses over the desired angular range for its use in the LWA. To demonstrate this, the proposed LWA is simulated in \texttt{ANSYS HFSS} by modeling the optimized Cell 1 and Cell 2 as ideal impedance boundaries as shown in Fig.~\ref{fig:impedance layer 1D lwa}(a).
%
%Here, to isolate the effects of angular sensitivity on radiation performance, all material losses are intentionally neglected. In particular, the refractive indices of Teflon and PI are set to 1.43 and 1.88, respectively. 
Here, to isolate the effects of angular sensitivity on radiation performance, all material losses are intentionally neglected, with the refractive indices of Teflon and PI set to 1.43 and 1.88, respectively. 

Fig.~\ref{fig:impedance layer 1D lwa}(b) shows the simulation results from which it is seen that the main beam is steered from $-27.6^\circ$ to $+22.6^\circ$ as the operating frequency increases from 1.9 THz to 2.6 THz. The simulated gain remains relatively constant across the scanning range, demonstrating robust frequency-dependent beam steering with minimal gain degradation. As a comparison, Fig.~\ref{fig:impedance layer 1D lwa}(c) shows a similar LWA configuration employing a conventional local metasurface, where each unit cell is designed at fixed incident angle of $60^\circ$. The main beam is steered from $6.2^\circ$ to $20.2^\circ$ as the operating frequency increases from 2.47 THz to 2.51 THz. A more realistic simulation incorporating physical unit-cell geometries and material dispersion and losses is discussed in the next section.

\section{Physical realization and Full-Wave Verifications}
\begin{figure}[t]
	\centering
	\resizebox{0.425\textwidth}{!}{\includegraphics{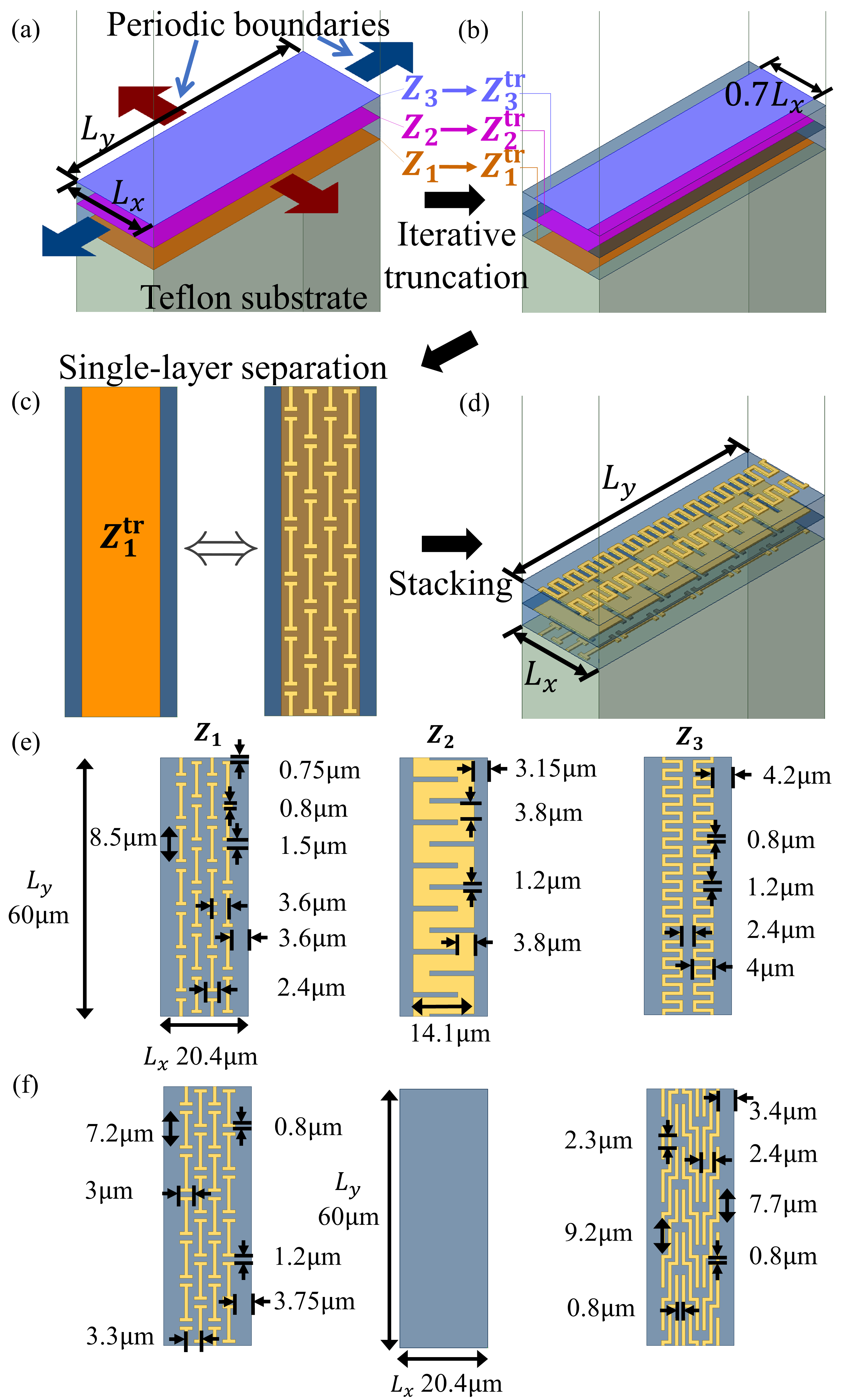}}
	\caption{Physical realization procedure of unit cells. (a) Ideal sheets in periodic boundary conditions. (b) Truncated impedance strips that possess similar angular responses. (c) A single layer separated from the multilayer stack and its equivalent physical pattern. (d) Complete unit cell constructed by stacking realized sheets. (e, f) Optimized geometries of the constituent impedance layers ($Z_1$, $Z_2$, and $Z_3$) arranged from left to right for (e) Cell 1 and (f) Cell 2. Note that the middle layer ($Z_2^{(2)}$) of Cell2 is implemented as a simple open circuit.}
	\label{fig:UC_LWA}
\vspace{-0.65cm}
\end{figure}

Figs.~\ref{fig:UC_LWA}(a-d) summarize the iterative truncation method utilized in this work to physically realize the proposed non-local THz metasurface.
The process begins by representing the optimized surface impedances as homogenized boundary conditions in \texttt{ANSYS HFSS} [Fig.~\ref{fig:UC_LWA}(a)]. While a layered stack can theoretically be realized by designing each impedance sheet independently, finite unit-cell periodicities introduce higher-order Floquet harmonics that couple neighboring layers and perturb the target angular response. Directly optimizing the full multilayer stack to account for these effects is computationally prohibitive due to the vast, interdependent parameter space.

To retain tractability while accounting for inter-layer interactions, we introduce an iterative truncation method [Fig.~\ref{fig:UC_LWA}(b)]. Each idealized sheet is incrementally truncated by 10\% along the $x$-axis, transforming it into an array of finite-width strips. After each truncation step, the surface impedances $Z_m^\text{tr}$ are fine-tuned to restore the optimized angular response, using the values from the previous iteration as a starting point. This process is repeated until the strips reach a final width of $0.7L_x$. This gradual refinement ensures that the desired non-local behavior is preserved even in the presence of strong inter-layer coupling. Based on this approach, the impedance values after truncation are determined to be $\{Z_1^\text{tr} = -j253.7,~Z_2^\text{tr} = j144.5,~Z_3^\text{tr} = j119.8\}~\Omega$ and $\{Z_1^\text{tr} = -j176.8,~Z_2^\text{tr} \xrightarrow{} \text{open circuited}, ~Z_3^\text{tr} = -j48\}~\Omega$.

\begin{figure}[t]
	\centering
	\resizebox{0.45\textwidth}{!}{\includegraphics{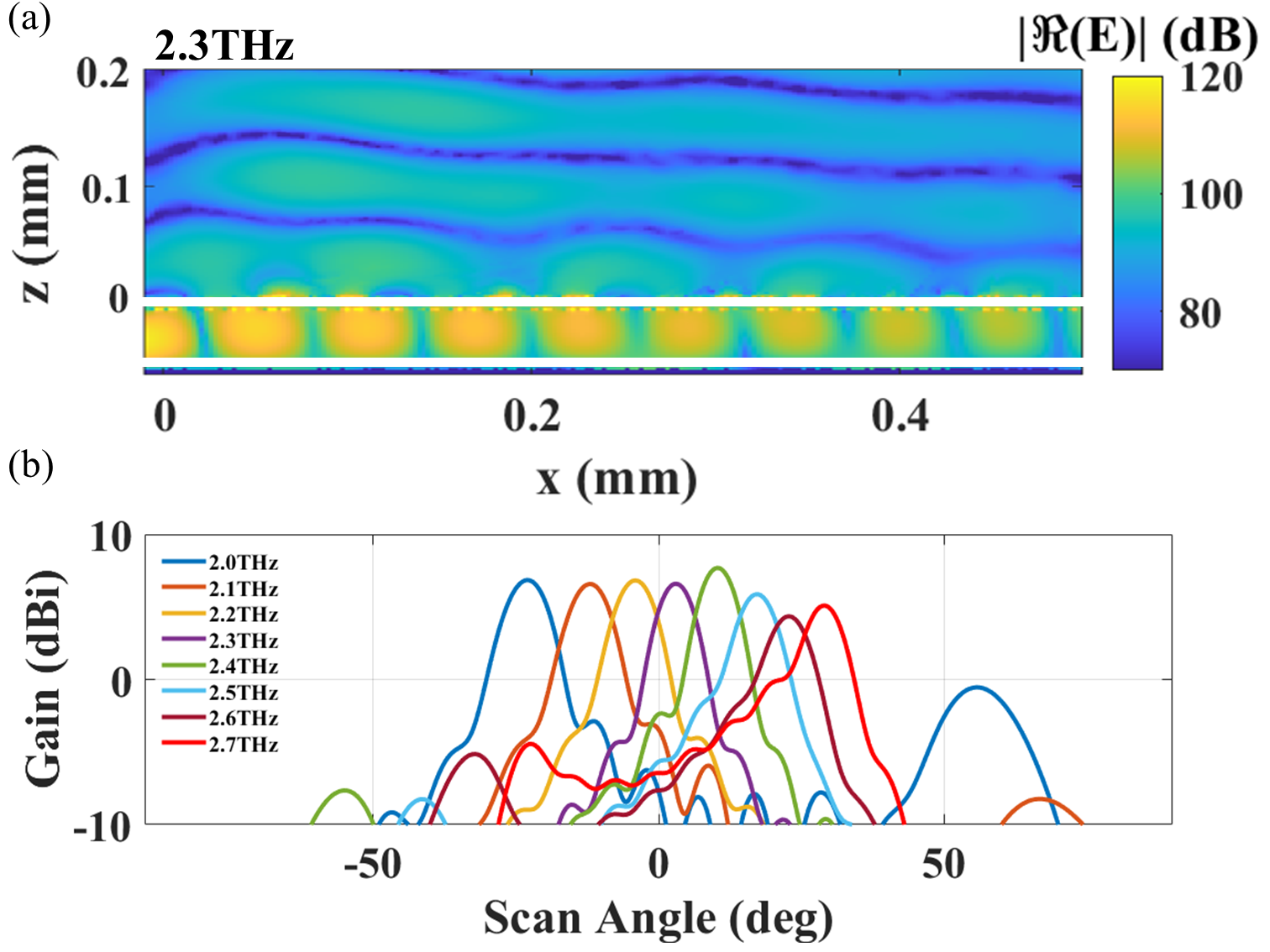}}
	\caption{Full-wave radiation performance of the physically
implemented angle-insensitive holographic LWA. (a)
Electric field distributions (dB scale) at 2.3 THz. (b) Realized gain spectrum
over the 2.0–2.7 THz band.}
	\label{fig:Physical HFSS result}
\vspace{-0.45cm}
\end{figure}

The physical implementations of Cell 1 and Cell 2 are then integrated into the LWA structure according to ~\eqref{eq:HoloPatt} to demonstrate the desired beam-steering capability. Full-wave simulations in \texttt{ANSYS HFSS} (Fig.~\ref{fig:Physical HFSS result}) demonstrate a stable beam-steering range from $-23.0^\circ$ to $+29.0^\circ$ over the 2.0–2.7 THz band.
Despite a minor 0.1 THz frequency shift due to Ohmic losses, the gain remains consistent across the 28\% fractional bandwidth. The side-lobe levels are maintained near or below $-10$ dB, confirming high beam purity. Notably, at 2.5 THz, the antenna achieves a peak gain of 5.87 dBi at $17.2^\circ$ (SLL of $-14.1$ dB), aligning closely with the $15^\circ$ holographic design target. These results validate the antenna's robust scanning performance and its ability to suppress the open-stopband effect.

\section{Conclusion}
In conclusion, we demonstrate a non-local metasurface-aided THz LWA that addresses the gain degradation inherent in local counterparts, where changes in the impinging angle alter the scattering response. The angular robustness is achieved by engineering four surface susceptibility parameters, for which we define theoretical performance limits and determine optimal values via numerical optimization. An iterative truncation method is introduced to accurately map these susceptibilities to physical unit cells. Full-wave validation in \texttt{ANSYS HFSS} demonstrates that the proposed LWA achieves a more than threefold extension in the scanning range compared to conventional local-metasurface designs.

\section*{ACKNOWLEDGEMENT}
This work was supported by the National Research Foundation of Korea (NRF) grants funded by the Korea government (MSIT) (RS-2024-00341191 and RS-2024-00343372).

\end{document}